\documentclass[letterpaper,floatfix,reprint,aps,pre,superscriptaddress]{revtex4-2}
\usepackage{graphicx}
\usepackage{gensymb}
\usepackage{tabularx}
\usepackage{color}
\usepackage{amsmath}
\usepackage{hyperref}

\newcounter{MD}
\newcommand*\MD%
{%
\ifnum\theMD>0%
{MD}\fi%
\ifnum\theMD=0%
{Molecular Dynamics (MD) \setcounter{MD}{1}}\fi%
}

\newcounter{SI}
\newcommand*\SI%
{%
\ifnum\theSI>0%
{SI}\fi%
\ifnum\theSI=0%
{Supplementary Information (SI) \cite{supp} \setcounter{SI}{1}}\fi%
}

\begin{document}

\preprint{APS/123-QED}


\title{Proofreading mechanism for colloidal self-assembly}

\author{Qian-Ze \surname{Zhu}}
\affiliation{School of Engineering and Applied Sciences, Harvard University, Cambridge MA 02139, USA}
\author{Chrisy Xiyu \surname{Du}}
 \affiliation{School of Engineering and Applied Sciences, Harvard University, Cambridge MA 02139, USA}
 \affiliation{Mechanical Engineering, University of Hawai`i at Mānoa, Honolulu HI 96822, USA}
\author{Ella M. \surname{King}}
 \affiliation{Department of Physics, Harvard University, Cambridge MA 02138, USA}
\author{Michael P. \surname{Brenner}}
 \affiliation{School of Engineering and Applied Sciences, Harvard University, Cambridge MA 02139, USA}
 \affiliation{Department of Physics, Harvard University, Cambridge MA 02138, USA}
\date{\today}

\begin{abstract}
    Designing components that can robustly self-assemble into  structures with biological complexity is a grand challenge for material science.  Proofreading and error correction is required to  improve assembly yield beyond equilibrium limits, using energy to avoid kinetic traps in the energy landscape. Here we introduce an explicit two staged proofreading scheme for patchy particle colloidal assemblies that substantially improves assembly yield and robustness. The first stage implements local rules whereby particles increase their binding strengths when they detect a local environment corresponding to a desired target. The second stage corrects remaining errors, adding a reverse pathway inspired by kinetic proofreading. The scheme shows significant yield improvements, eliminating kinetic traps,  giving a much broader temperature range with high yield. Additionally, the scheme is robust against quenched disorder in the components. Our findings illuminate a pathway for advancing programmable design of synthetic living materials, potentially fostering the synthesis of novel biological materials and functional behaviors.
\end{abstract}

\maketitle

Living matter exhibits far more complicated structures and dynamics than those we create synthetically. For example, enzymes exquisitely regulate metabolic pathways \cite{newsholme1973regulation}; microtubules and virus shells can self-assemble and disassemble on cue \cite{knossow2020mechanism, vanburen2005mechanochemical, bruinsma2021physics, buzon2021virus}; in DNA proofreading, polymerase massively reduces binding errors beyond equilibrium yield \cite{reha2010dna}. Inspired by living matter, a longstanding goal is to create synthetic materials that exhibit this level of biological complexity. An essential component of a strategy is using programmable building blocks, with examples ranging from DNA nanotechnology\cite{iinuma2014polyhedra,ke2009multilayer,rothemund2006folding}, to biological assemblies like clathrin, where triskelia are programmed to interact for emergent behaviors \cite{morris2019cryo}. In all of these cases, building blocks are programmed to achieved properties that far exceed the functionality of a random set. 
%
Yet, on its own, programmability has limits \cite{zeravcic2014size,murugan2015undesired,huntley2016information}. Even with programmable components, there are strict limits to complexity and size of the emergent structure. Programmable models in DNA origami create complicated structures yet with uncontrollable errors \cite{rothemund2006folding}. 

\begin{figure}
    \centering
    \includegraphics[width = 1.0\linewidth]{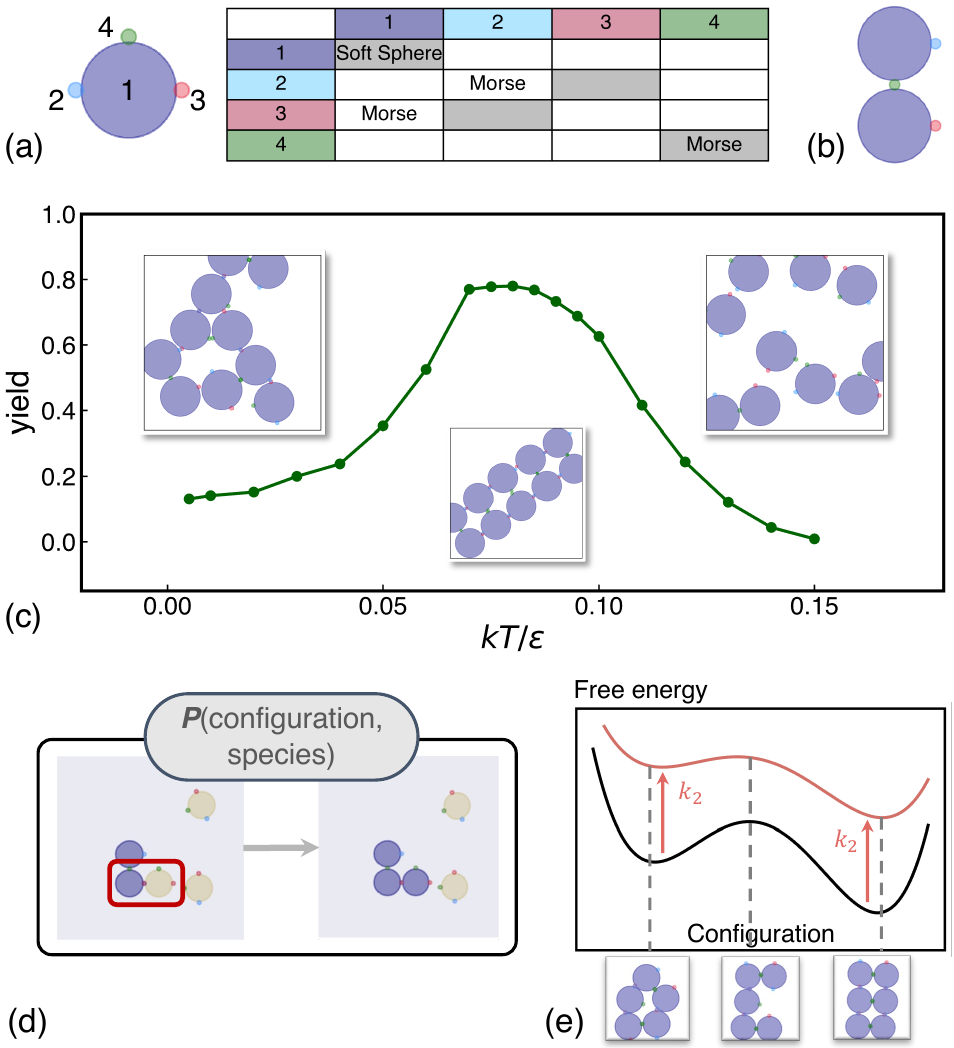}
    \caption{\textbf{System and two-stage proofreading mechanism} (a) Left: Individual patchy particle consisting of a central sphere (1: purple) and three distinct patches (2: blue; 3: red; 4: green). Right: Interaction matrix illustrating inter-component interactions. Grey squares represent soft sphere or Morse potential as specified. Blank squares denote no interactions. (b) Schematic illustration of the seed particle. (c) Yield curve for a conventional mono-species system. x axis is the temperature $k_BT$ rescaled by the interaction potential energy $\epsilon$. Inset: exemplary final configurations of 40000-step forward simulations in (left) kinetic regime, (middle) optimal regime
    (right) entropically-favored monomer regime.
    (d) Illustration of the state change mechanism. Yellow represents weaker interactions, and purple represents stronger interactions. Only correct bonds (highlighted by the red box) are reinforced. (e) Illustration of the full proofreading mechanism via the engineer of free energy landscape. Insets display (left) a local defect; (middle) configuration at the energy barrier; (right) part of a correctly assembled structure.}
    \label{fig:1}
\end{figure}

To create more robust assembly pathways, we hypothesized that it is necessary to proofread errors during assembly pathways. To explore this, we aimed to construct the simplest system where proofreading could be implemented. Given that enzymatic activity requires bond directionality, isotropic colloids \cite{meng2010free,zeravcic2017colloquium} seem insufficient, so we used a set of identical patchy particles (see Fig.~\ref{fig:1}(a)) in two spatial dimensions, powerful enough to capture the complex properties yet simple enough to design \cite{pawar2010fabrication}. Each patch can be programmed to interact with other patches, by modifying its interaction strength and range.
Our model for proofreading is inspired by the dynamic growth of microtubule \cite{knossow2020mechanism, vanburen2005mechanochemical} and allosteric interactions in protein binding \cite{nussinov2014principles, ni2019allosteric}, where bonds can be enzymatically catalyzed to strengthen or weaken when their local environment assembles correctly. To capture this, we ascribe to each particle the concept of a {\sl state}, which dictates how the patches on a particle interact with other particles. Each particle has an identical initial state; yet as assembly proceeds, particles are able to measure their local neighborhood and correspondingly change their state. A particle with changed state has patches with different interaction strengths, ranges and (potentially)  locations. Our key question is whether it is possible to program state changes to achieve the assembly of particular structures. Note that state change implies energy consumption: the energy an individual particle contributes to the configuration energy changes when it changes state. State change is a programmed allosteric response, a synthetic analogue of allostery \cite{changeux2012allostery}.

The design space for programmable patchy particles with state change is enormous: for a given structure, we aim to find a series of transitions of patches that optimize assembly yield.  To find the optimal parameters, we leverage automatic differentiation methods\cite{baydin2018automatic}, which made it possible for  modern machine learning to search models with billions of parameters. This allows us  to navigate the parameter space of a complex model with gradient descent. Previous research has demonstrated that this method can efficiently find patch locations and strengths for designing structures with patchy particles, as well as discover new physics \cite{jaxmd_patchy}. 

Specifically, inspired by microtubule growth, we aim to develop rules to robustly grow a two-dimensional ladder-like structure. 
The individual patchy particle consists of a central particle that interacts via a soft sphere potential and three patches that interact via a Morse potential, see the illustration and interaction matrix in Fig.~\ref{fig:1}(a).
All subsequent studies were conducted using molecular dynamics simulation under constant temperature and volume condition, and with periodic boundary condition (see \SI for more simulation details). 
We begin the assembly with a seed consisting of two rigidly connected particles, with valid contacts only in one direction (Fig.~\ref{fig:1}(b)).
Fig. ~\ref{fig:1}(c) shows the yield curve for assembly of this structure, quantified in a system with 10 particles.
The yield is defined as the ratio of the number of instances of error-free assemblies with size 10 to the total number of instances (for a detailed definition see \SI).
In the low temperature regime (left), the structures are kinetically trapped in local minima of free energy, while in the high temperature regime (right), dissociated monomers are entropically favored over stable bonds. 
The maximum yield is around 90 percent, a high number, but perfect assembly does not occur even in this extremely simple structure. Our goal is to develop a proofreading scheme to improve suboptimal yields.

%
%
 
%
%
%

We design the proofreading scheme as follows: errors in assembly occur because of local minima, whereby particles form a ring with a smaller number of contacts than the dense ladder (see the left inset of Fig.~\ref{fig:1}(e)). To inhibit this, we need a rule where a particle only strengthens its bonds when its {\sl local environment} meets desired criteria. This process is depicted in Fig.~\ref{fig:1}(d): the initial state of the particle (yellow, species 1) has  weak interactions $E_1$. When the right bonds are made in its local environment (red box in Fig.~\ref{fig:1}(d)), it changes state (purple, species 2) strengthening the  interactions to $E_2$. This is the first stage of the proofreading process. Errors still occur after the first stage because of mismatches with different local environments. We therefore introduce a second stage, inspired by kinetic proofreading \cite{hopfield1974kinetic}, whereby we add a reverse pathway to decrease the bond strength of randomly chosen particles. This modulates the free energy landscape to favor the transition from local minima to the target structure (Fig.~\ref{fig:1}(e)).
%

%

%
We demonstrate in what follows that when the rules are properly chosen, this two-stage proofreading mechanism substantially increases the yield of the desired structures over a much broader temperature range. 

\begin{figure}
    \centering
    \includegraphics[width = 1.0\linewidth]{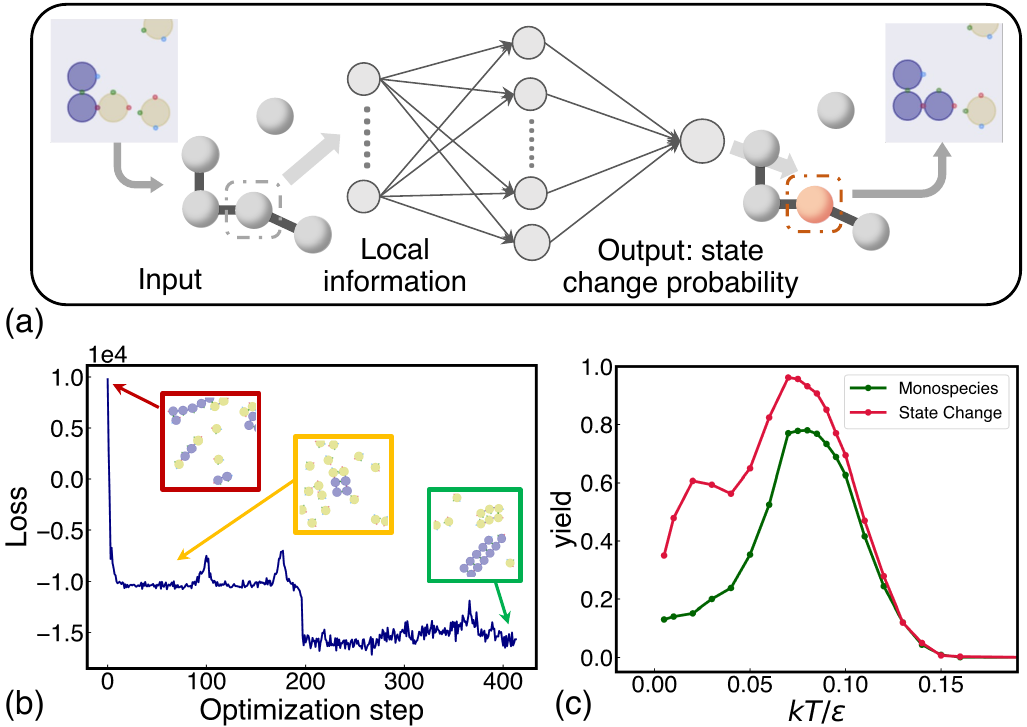}
    \caption{\textbf{State change} (a) Schematic illustration of GNN as the update rule for state change. (b) The loss trajectory during GNN training. 
    Inset: 100000-step forward simulation using the indicated parameters at optimization step 0 (red), 50 (yellow), and 400 (green). (c) Yield curves for (green) conventional mono-species and (red) state change systems. $x$-axis is temperature rescaled by the dominant interaction energy scale. Note for the state change system, the interaction energy scale refers to the strong bond strength $E_2$.}
    \label{fig:2}
\end{figure}

\textit{State change.} -- In this first stage of our proofreading scheme, we program a local update rule for each particle that senses its surrounding environment, and determines whether it matches the desired target. If so, the particle changes state, strengthening its bonds with its neighbors. 
The key question is to figure out how to program the rule for changing state: if the rule is either too permissive or too stringent in defining the desired local environment, assembly yield is degraded (See \SI). The space of possible local rules is large: to search it systematically, we parameterize the 
 sensing and decision making process with a graph neural network (GNN), which allows us to easily encode the geometry of the local environment
 \cite{battaglia2018relational}. 
Given the local environment of the particle within a prespecified cut off range, the network predicts the probability that the particle changes state. 
We learn the parameters of the GNN to improve the yield of our target structure, using automatic differentiation to back-propagate the gradient of a loss returned from a batch of forward molecular dynamics simulations. 
Specifically, we construct the GNN (Fig.~\ref{fig:2}(a)) as follows \cite{bapst2020unveiling}: 
First, we encode the local environment of a particle into the node and edge embedding of the graph. The node encodes the states of neighboring particles, whereas the edge includes pairwise distances between particle centers and patches. We choose a cutoff range of three times the radius, covering nearest neighbors and next nearest neighbors in the target structure. 
The GNN then aggregates the local information from the neighbors of each particle and updates its edge and node embeddings successively using multi-layer perceptrons (MLP). Ultimately this information is decoded to give the probability of state change of each particle. The MLPs have 79 parameters in total, which we denoted by 
$\hat{\theta}$.
To optimize these 79 parameters simultaneously in the context of a molecular dynamics simulation of assembly, we use JAX-MD,  an end-to-end differentiable and GPU-accelerated MD engine \cite{jaxmd_git}. To train the model, we measure the performance of the update rule with a loss function indicating how far the final structure maps to the desired one:
$\mathcal{L} = \mathcal{L}(s_1, s_2, \cdots, s_N; \hat{\theta})$, where $s_i$ represents the state of particle $i$, which includes position, orientation, species and other information.
The loss function has two terms (see \SI), one penalizing deviations from our desired target (the compact ladder structure), with the other encouraging elongation of the structure .

Since state change is an intrinsically discrete process, to ensure differentiability, we adapt the REINFORCE algorithm (a.k.a. score function) \cite{williams1992simple}, making it possible to take gradients through the simulation. 
The probability that particle $i$ goes through state change is a function of its current state and the GNN parameters $\rho(s_i(t); \hat{\theta})$. 
Therefore the objective of the optimization, an ensemble average of the loss function at the final simulation step $t$, can be written as,
\begin{equation}
    \begin{aligned}
    \langle \mathcal{L}\rangle = \int_{\substack{\text{configuration} \\ \text{space}}}
      &\mathbf{\mathcal{D}s}(t) \mathcal{L}(s_1(t),\cdots,s_N(t); \hat{\theta}) \\  
      &\cdot P[s_1(t),\cdots,s_N(t); \hat{\theta}]
    \end{aligned}
\end{equation}
where $P[s_1(t),\cdots,s_N(t); \hat{\theta}]$ is the joint probability of the final configuration of the whole system $[s_1(t),\cdots,s_N(t)]$ with the current parameters $\hat{\theta}$,
\begin{equation}
    P[s_1(t),\cdots,s_N(t); \hat{\theta}] = \prod_{\tau = 0}^t \prod_{i = 1}^{N}\rho(s_i(\tau); \hat{\theta})
\end{equation}
Using the REINFORCE algorithm, we can take gradients of the loss function with respect to the parameter $\hat{\theta}$,
\begin{equation}
    \begin{aligned}
        \nabla_{\hat{\theta}}\langle \mathcal{L}\rangle & = \int \mathbf{\mathcal{D}s}(t) \nabla_{\hat{\theta}}( \mathcal{L}(\mathbf{s}(t); \hat{\theta}) P[\mathbf{s}(t), \hat{\theta}])\\
        & = \int \mathbf{\mathcal{D}s}(t) P(\nabla_{\hat{\theta}}\log P)\mathcal{L} + \int \mathbf{\mathcal{D}s}(t) P\nabla_{\hat{\theta}}\mathcal{L} \\
        & = \langle (\nabla_{\hat{\theta}}\log P)\mathcal{L} \rangle + \langle \nabla_{\hat{\theta}}\mathcal{L} \rangle
    \end{aligned}
\end{equation}
where the first term is the REINFORCE gradient, and the second term is the conventional gradient.

To optimize the state change parameters, we consider a system consisting of 21 particles at temperature $k_BT = 1.0$, with weak interaction $E_1 = 4.0$, strong interaction $E_2 = 50.0$. 
Following the optimization procedure described in \cite{jaxmd_patchy}, we use Adam optimizer to run 400 optimization steps with a concatenated learning rate scheme of $(0.005, 0.001)$. 
Each optimization step is performed by running 512 replicates of 41000-step forward simulation and back-propagating the loss over the last 1000 steps of the simulation. 
This makes it possible to store the trajectories in memory on an A100 with 80GB of RAM.  
We include a python notebook demonstrating the basic construction of the mechanism and optimization of the GNN update rule
\cite{notebook_git}.

We find that it is difficult for the learning rule to converge when we initialize the GNN with random initial parameters. Instead we pretrain the model by initializing it with logical rules for the desired local structure (see \SI), which allows  optimization to rapidly converge to a set of optimal parameters, as shown in Fig.~\ref{fig:2}(b). 
The insets show the final configurations of 100000-step forward simulations using the GNN parameters at different optimization stages.
Curiously, the training dynamics exhibits an initial plateau in the loss trajectory before converging to a lower state. 
This indicates that the system is initially trapped in a local minimum of the optimization, where it tries to minimize the deviation from ideal ladder structure as much as possible. 
This consequently inhibits the growth, as indicated in the yellow inset of Fig.~\ref{fig:2}(b). 
It then makes a few attempts to jump out of the local minimum, marked by the intermittent spikes.
Ultimately, the system successfully learns the optimal growth rule balancing mitigating errors with growth. 

Fig.\ref{fig:2}(c) shows the yield curve for the state change scheme compared with the mono-species assembly. For a fair comparison with Fig.~\ref{fig:1}(c), we rescale the temperature with respect to the strong interaction energy $E_2$ after state change.
The state change scheme greatly improves the yield, especially in the regime where structures are typically kinetically trapped in local minima. 

\begin{figure}
    \centering
    \includegraphics[width = 1.0\linewidth]{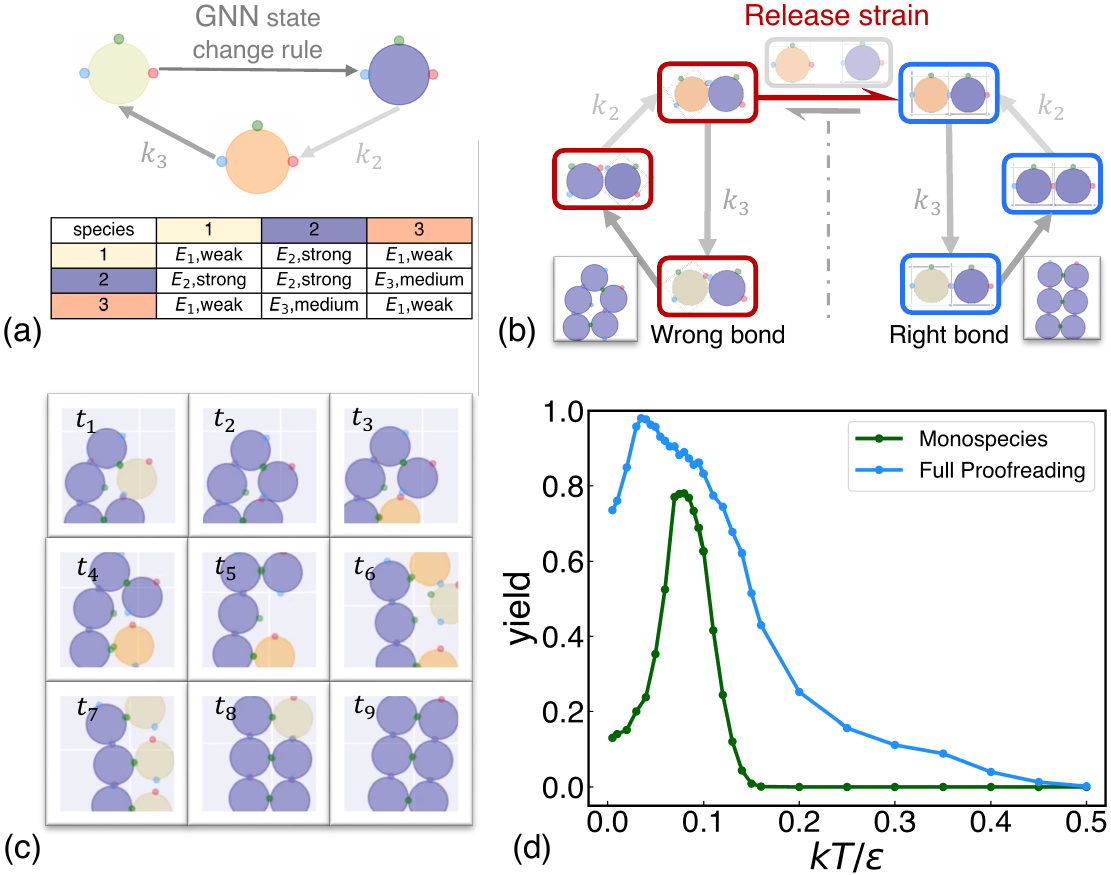}
    \caption{\textbf{Full proofreading mechanism} (a) Schematic illustration of the full proofreading mechanism, including both the state change and the reverse pathway. Lower panel: interaction matrix between the three particle species. (b) Step-by-step illustration of the proofreading mechanism. (c) Snapshots of the proofreading dynamics in an increasing-time order. (d) Yield curve for (green) conventional mono-species and (blue) full proofreading mechanism. $x$-axis is temperature rescaled by the dominant interaction energy scale. Note for proofreading mechanism, the interaction energy scale refers to the medium bond strength $E_3$.}
    \label{fig:3}
\end{figure}

\textit{Reverse pathway.} -- Yet, the state change scheme still leaves errors, where the local environments around different particles are not registered with each other.  When state change occurs, the bonds essentially become irreversible, so any errors are frozen in (for illustration see \SI).  We find that there are mainly two kinds of errors that the state change rule produces: Either free floating particles accidentally undergo state change, or incorrectly bonded particles undergo state change resulting in dislocation in the structure (Fig.~\ref{fig:3}(b) left panel).

The second stage of our proofreading scheme aims to fix these issues by adding  a reverse pathway to weaken bonds. Inspired by kinetic proofreading \cite{hopfield1974kinetic}, we randomly choose particles to transition to an intermediate species 3 with moderate bond strength $E_3 = 15.0$. 
The transition dynamics and interaction matrix between these three species is shown in Fig.~\ref{fig:3}(a). 
We choose the probability of a species 2 (purple) particle to transition into species 3 (orange) particle to be $k_2 = 2\times 10^{-5}$, and the probability of a species 3 (orange) particle to transition into species 1 (yellow) particle is $k_3 = 2\times 10^{-4}$.
This reverse process is identical for all particles in the structure, and does not depend on whether the bonding environment is correct.

We reason that if a particle's bonding environment is correct, it will quickly reform its original structure through the state change mechanism (the right loop of Fig.~\ref{fig:3}(b)).
If the bonding environment is in a local minimum, it has a chance to correct itself, breaking an incorrect bond, and ultimately stabilize in the correct structure through the state change mechanism, as shown in Fig.~\ref{fig:3}(b). 
Snapshots in Fig.~\ref{fig:3}(c) and the accompanying supplementary videos clearly show the error correcting process.
The reverse pathway effectively modulates the free energy landscape,
elevating the activation barrier for escaping local minima to a level that is comparable to the thermal energy ($\Delta E_{LM} \lesssim k_B T$).
While the energy depth of the ground state remains significantly greater than the thermal energy ($\Delta E_{GS} \gg k_B T$), as illustrated in Fig.~\ref{fig:1}(c). 
This results in a net transition from the wrong bonds to the right bonds, thereby substantially improving the yield.
%
%

\begin{figure}
    \centering
    \includegraphics[width = 1.0\linewidth]{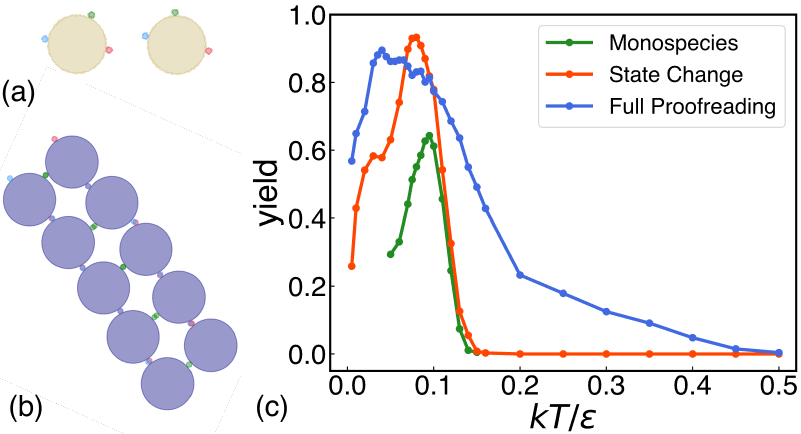}
    \caption{\textbf{Quenched disorder} (a) Examples of individual particles with quenched disorder. (b) A correctly assembled structure with quenched disorder. (c) Yield curve for (green) conventional mono-species, (red) state change mechanism and (blue) full proofreading mechanism. $x$-axis is temperature rescaled by the dominant interaction energy scales.}
    \label{fig:4}
\end{figure}


Fig.\ref{fig:3}(d) shows the result for yield as a function of rescaled temperature for the full two-stage proofreading scheme compared with the conventional mono-species assembly. 
For a fair comparison, we again rescale the temperature with respect to the dominant energy scale, which is the intermediate interaction $E_3$ for proofreading.
The proofreading scheme not only greatly improves the yield, but also largely broadens the optimal temperature range where the yield is relatively high, both in the kinetic and thermal regimes. 

\textit{Quenched disorder} -- In addition to thermal noises, there may also be intrinsic disorder in the geometry of the patches, whereby the patches do not form at exactly the optimal configuration. Is our proofreading scheme robust to such errors? To model this quenched disorder, we perturb the location of the patches in individual components away from their optimal ($90^\circ$ angle) geometry. For every  patch ($\alpha = 1,2,3$) on every particle ($i=1,\cdots, N$) we move the angle by $\varepsilon_{i\alpha}$ from its regular location, where $\left\{\varepsilon_{i\alpha}\right\}$ are drawn from a uniform distribution bounded by the magnitude of the quenched disorder. 
Note that in this scheme every particle in the simulation is manufactured slightly differently. 
Here the magnitude of disorder is set to be 0.2 (radian). Fig.~\ref{fig:4}(a) illustrates two examples of such disordered particles. Fig.~\ref{fig:4}(b) shows a correctly assembled structure using these disordered particles. 
%
However, the disorder makes it substantially less likely to form error-free structures because defects will accumulate in the assembly process.

We use the identical proofreading scheme described above, without retraining the parameters and  again calculate the yield curves for each of the three cases (conventional mono-species, state change, full two-stage proofreading) with the disordered components. 
To fairly compare the yield with intrinsic disorder in the system, we loosen the threshold in yield definition (see \SI). Fig.\ref{fig:4}(c) shows the yield for the conventional mono-species assembly (green curve) decreases substantially with quenched disorder. Both the state change (red curve) and the full proofreading (blue curve) recover high yields, demonstrating the robustness of the proofreading mechanism.

\textit{Discussion.} --
Here we have demonstrated that it is possible to program proofreading mechanisms to substantially improve  yield during self assembly. Although we have studied a particular example of colloidal assembly, the principles uncovered here could have much broader applicability.
The basic idea of the proofreading scheme is conceptually simple: every component is able to validate its local environment, and when the environment is correct, it undergoes a state change to increase bond strengths. Such allosteric changes occur in biology, catalyzed by hydrolysis of an energy containing molecule like GTP orATP. The state change mechanism requires substantial tuning to work most effectively: we accomplished this by parameterizing the decision rule with a neural network and then directly optimizing it through a molecular dynamics simulation of the assembly process. 
The resulting scheme with state change is highly efficient but still suffers errors. This is predominantly because defects can still arise due to discordances between local environments. To fix these, we introduce a proofreading scheme inspired by Hopfield's kinetic proofreading\cite{hopfield1974kinetic}. Particles that have undergone state change randomly transition to states with weaker bond strengths which allow these errors to be corrected. From this configuration, the probability of correcting an error is much higher than the probability of forming a new error, and hence this results in nearly perfect yield.

While the conceptual foundations of these ideas has long existed \cite{changeux2012allostery,hopfield1974kinetic}
to our knowledge this is the first time that this has been explicitly programmed in a simulation of colloidal particles. Although our demonstration used a simple structure to assemble, both the concepts and the programming framework we outline here are general and should work for assembly of other structures with other types of components. Goals for future work should apply these schemes to other structures such as the assembly of 3D shells that can encapsulate or release their contents, or to other functional behaviors, such as self-replication and disassembly. 

A key aspect of this work is the use of differentiable molecular dynamics simulations with patchy particles, a possibility enabled by the development of patchy particle simulations within a differentiable library \cite{jaxmd_patchy}. The ability to carry out this type of optimization within a molecular dynamics simulations opens up enormous possibilities for the design of novel components. Furthermore, the ability to program interactions and state change opens the door to even more complicated  structures and behaviors. As examples, we could design components to engineer the yield shape itself, by optimizing the width and temperatures where particular structures occur. Alternatively, the system could be programmed to exhibit multifarious assembly \cite{murugan2015multifarious}, whereby different seeds nucleate different structures. Yet another direction is to consider the distribution of manufacturable components and directly optimize over this distribution, for example optimizing the distribution of the position of patches. To facilitate the development of these possibilities, we have have open sourced the code coupling molecular dynamics simulations to the training of local rules \cite{notebook_git}.

This work is the first step to a long standing goal of building synthetic materials with the properties of living ones \cite{zeravcic2017colloquium}. Being able to design these materials in a computer is a necessary but not sufficient condition for implementing them in experiments. We now must develop these computational approaches in systems that can be experimentally realized, and tune the computational models to reflect the experimental system accurately enough that we can use computation to guide experiments. The development of such a system is an important and unsolved problem. Possible experimental systems range from magnetic handshake materials \cite{niu2019magnetic}, to DNA origami \cite{ke2009multilayer, rothemund2006folding}.

\textbf{Acknowledgments:} We thank Francesco Mottes for suggesting that we consider quenched disorder, and all members of the Brenner group for a stimulating environment to use automatic differentiation technologies for optimizing dynamical problems. This material is based upon work supported by the Office of Naval Research (ONR N00014-17-1-3029, ONR N00014-23-1-2654),  the NSF Grant DMR-1921619, and the NSF AI Institute of Dynamic Systems (\#2112085).

\bibliographystyle{apsrev4-2}
\bibliography{main}
\end{document}



\title{Supplemental Material for:\\ Proofreading mechanism for colloidal self-assembly}

\author{Qian-Ze \surname{Zhu}}
\affiliation{School of Engineering and Applied Sciences, Harvard University, Cambridge MA 02139, USA}
\author{Chrisy Xiyu \surname{Du}}
 \affiliation{School of Engineering and Applied Sciences, Harvard University, Cambridge MA 02139, USA}
 \affiliation{Mechanical Engineering, University of Hawai`i at Mānoa, Honolulu HI 96822, USA}
\author{Ella M. \surname{King}}
 \affiliation{Department of Physics, Harvard University, Cambridge MA 02138, USA}
\author{Michael P. \surname{Brenner}}
 \affiliation{School of Engineering and Applied Sciences, Harvard University, Cambridge MA 02139, USA}
 \affiliation{Department of Physics, Harvard University, Cambridge MA 02138, USA}
\date{\today}

\maketitle

\section{Simulation details}
In this section, we provide the specific parameters and methods employed in our simulations. 
We conduct molecular dynamics simulation under constant temperature and volume condition with periodic boundary condition. Specifically we use the Nosé–Hoover thermostat to carry out the simulation. 
When training the graph neural network (GNN) as the state change update rule, the simulation is kept at temperature $k_BT = 1.0$, and the size of the system is chosen to be 1 seed (can be regarded as two particles), and 19 normal particles, with a volume fraction of $0.2$. 
When evaluating the yield curve, the temperature is swept from $0.05 E$ to $0.5 E$, where $E$ is the dominant energy scale, and the system consists of 1 seed and 8 normal particles, with a volume fraction of $0.4$. 
The simulation timestep is set to be $dt = 10^{-3}$.

For each individual patchy particle which is considered as rigid body, the central particle has a radius of 1.0, and patches are treated as points, i.e. sizeless.
The central particles interact via soft sphere repulsion and patches interact via Morse potential. 
The specific expressions for these two interaction potentials are as follows:
\begin{itemize}
\item[] Morse Potential:
\begin{equation}\label{eq:morse}
    U(r) = \varepsilon (1 - e^{-\alpha (r - r_0)})^2
\end{equation}
\item[] Soft Sphere Potential:
\begin{equation}\label{eq:soft}
    U(r) = \varepsilon\left(\frac{\sigma}{r}\right)^\alpha
\end{equation}
\end{itemize}
The parameters for the Morse potential are set to be: interaction range $\alpha = 5.0$, $r_0 = 0.0$ (i.e. the patches are sizeless), interaction strength for species 1 (weak interactions) $\varepsilon_1 = 4.0$, for species 2 (strong interactions) $\varepsilon_2 = 50.0$, for species 3 (medium interactions) $\varepsilon_3 = 15.0$. 
The parameters for the soft sphere potential are set to be: interaction range $\alpha = 2.0$, repulsion strength $\varepsilon = 10000.0$, radius $r = 1.0$.


\section{Final Configuration and Yield Curve}
For consistency with the conventional definition of the yield, here we consider a system comprises a fixed particle count of 1 (seed) and 8 (normal particles). Yield is counted as 1 only if all particles assemble into the desired compact ladder structure without local defects. Any deviation from this precise arrangement results in the yield being quantified as 0. 

\subsection{Final configuration}
\label{final_config}
\begin{figure}
    \centering
    \includegraphics[width = 0.5\linewidth]{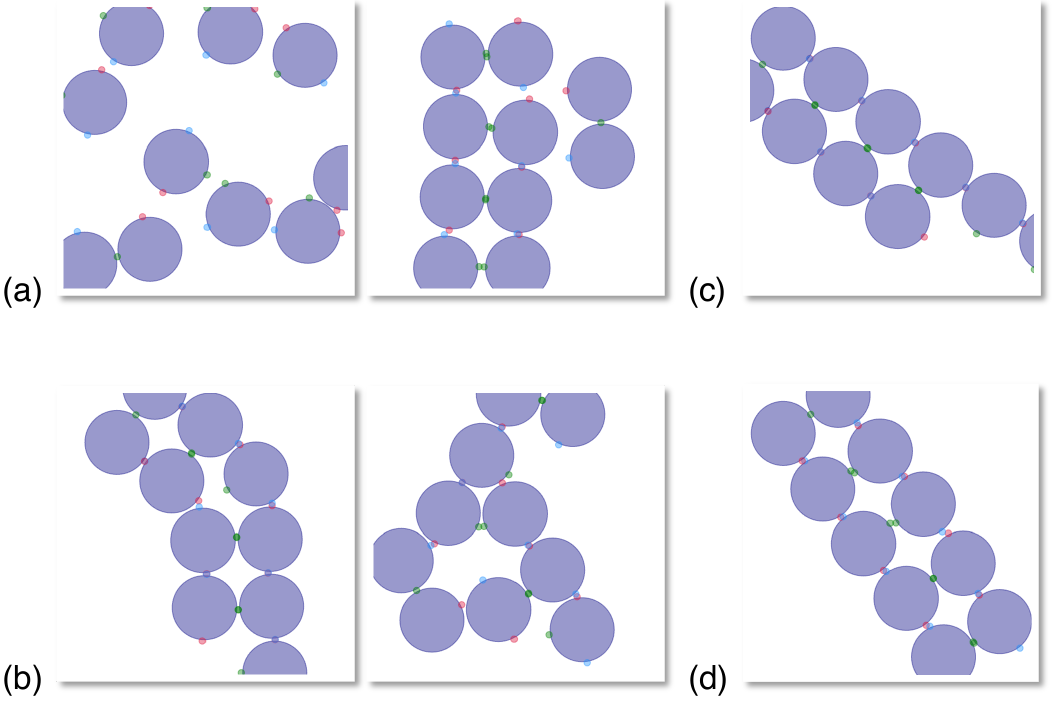}
    \caption{(a) Case I, incomplete structure. Left: free floating normal particles; right: unbound seed. (b) Case II, structures with local defects. (c) Case III, compact ladder structure but with unequal arms. (d) Case IV, perfect compact ladder structure.}
    \label{fig:s1}
\end{figure}

We closely examined the final configurations of all simulation results, and divide them into four different kinds of structures (see corresponding examples in Fig.~\ref{fig:s1}):
\begin{itemize}
\item[I.] Incomplete structures, with free particle floating around, including unbound seeds. The latter case, which is a typical bulk defect in dense systems, is a primary factor contributing to low yields in the conventional mono-species assembly.
\item[II.] Structures with local defects, for example, the dislocation in the local minimum emphasized in the main text.
\item[III.] Compact ladder structure but with unequal arms.
\item[IV.] Perfect compact ladder structure with equal arms.
\end{itemize} 

For these four kinds of configurations, we considered the last two (i.e. case III and case IV as correct structures. As in Case III, there is no local defect and will grow into the perfect ladder if more individual components are provided.

\subsection{Formal definition of the yield}
\label{defyield}
The definition of the yield does not need to be differentiable as in the case of loss function. But the yield needs to be precise to distinguish between the desired structure and those with defects.

For the perfect compact ladder structure (Case IV), we consider two metrics below:

\begin{itemize}
\item[1).] The distance and orientation difference between each particle compared to the exact diagonal neighbor. This criterion is aiming for discriminating against those cases with local defects, where there are usually vacancies or dislocations in the place of a perfect diagonal neighbor.

For each particle, we find the neighboring particle that has the pairwise distance closest to $2\sqrt{2}r$. Then we compare the distance and orientation difference of these neighbors to the perfect diagonal particle, which should have distance of $2\sqrt{2}r$ and orientation difference of $\pi$. If these two differences are within certain thresholds, then this particle passes the first test. The threshold for the distance difference is set to be $c_1 = 0.22$, while the threshold for the orientation difference is set to be $c_2 = 0.35$. These values are determined after examining the final configurations of what are regarded as correct versus incorrect cases. 
Additionally, we swept these thresholds across a range of different values (see Section~\ref{sweep}), and the primary conclusion in the main text still holds.

\item[2).] The distance between patches. This criterion serves to check whether the particle is in direct contact with other particles. This can discriminate against those incomplete cases with free floating particles that are not connected with any others. 

For each particle, we only consider its patch 1 and 2 (defined in the main text).
This choice is intended to exclude out-of-order growth which tends to end up as local defects. 
For each patch 1, we find the closest patch 2 and vice versa.
If the distance between a pair of these patches falls below a specified threshold, the particle is considered to be directly connected to another particle.
The threshold for determining direct contact between patches is set to be $c_3 = 0.2$. Similar to the previous metric, different threshold values were analyzed in Section~\ref{sweep}.
\end{itemize}

Combining these two criteria, if both are satisfied, then the particle is in the correct local environment corresponding to the last case (case IV) in Section~\ref{final_config}. 
Otherwise we should check if this particle belong to Case III, which is the special case of unequal arms.
In this case, we only need to check whether the particle is at the end of the dangling tail, as particles located in the middle exhibit the same local environment as in case IV. 
Therefore we further check whether only one of the particle's patch 1 and 2 is in direct contact with the patch 2 and 1 of any other particles, and in the meantime its patch 3 is not in direct contact with any other patches (in order to exclude out-of-order growth). The threshold $c_3$ for determining direct contact between patches is the same as above.

Specifically, the exact values of the thresholds are determined by sweeping $c_1$ from $[0.2, 0.22, 0.25]$, $c_2$ from $[0.2, 0.35, 0.5]$, $c_3$ from $[0.1, 0.2, 0.3]$, and looking at the corresponding configurations and deciding which values can best reflect the performance.

To conclude, we check if every particle satisfies case IV or III. As long as one particle does not satisfy either of these two cases, the whole instance is counted as an incorrect structure with zero yield.
The code for implementing this yield function is provided in the accompanying python notebook \cite{notebook_git}.

To ensure equilibrium, we record the yield trajectory of a $10^7$-step forward simulation, where the yield turns out to reach a constant level after $2\times 10^6$ steps. Consequently, we choose the simulation steps to be $5\times 10^6$ for evaluating the yield.

\subsection{Sweeping parameters in the yield definition}
\label{sweep}
In this section, we analyze how the threshold values in the yield definition affect the yield curves under three different mechanisms, i.e. i). the conventional mono-species system; ii). the state change mechanism; iii). the full two-stage proofreading mechanism.

\begin{figure}
    \centering
    \includegraphics[width = 1.0\linewidth]{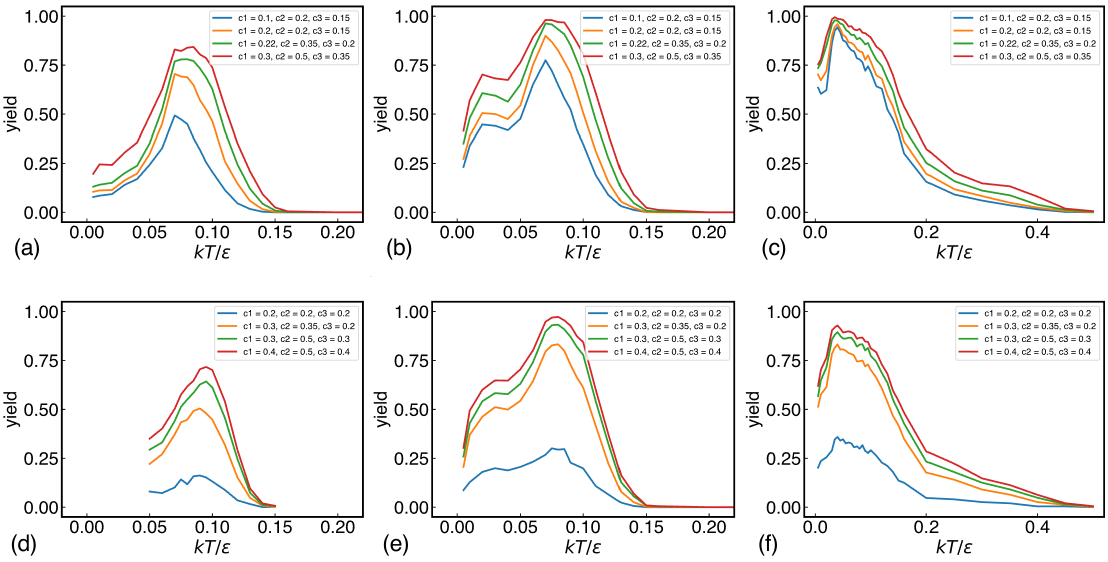}
    \caption{Yield curves with threshold values noted in the legend, for (a,d) conventional mono-species system, (b,e) state change mechanism, (c,f) full two-stage proofreading mechanism, with ideal component. (d-f) correspond to the cases with quenched disorder.}
    \label{fig:s2}
\end{figure}

Here we present four yield curves corresponding to distinct sets of representative threshold values in Fig.~\ref{fig:s2} (a-c). The results demonstrate that although the thresholds indeed affect the quantitative values of the yields, the overarching trend remains consistent, i.e. the proofreading mechanisms indeed greatly outperform the conventional mono-species assembly, as well as maintain a much wider optimal temperature range.
Consequently, the primary conclusion of this study is not affected by variations in the threshold values.

\subsection{Sweeping yield parameters with quenched disorder}
For fair comparison, in scenarios with quenched disorder, we should adjust the thresholds in the yield definition to account for the intrinsic disorder in the individual components.
Specifically, for the yield curve in Fig. 4 of the main text, we set the thresholds to be $c_1 = 0.3$, $c_2 = 0.5$, $c_3 = 0.3$. 
Similarly we also swept the threshold values and demonstrated the robustness of the conclusions. The results are shown in Fig.~\ref{fig:s2} (d-f) respectively.


\section{Hand-designed update rule}
For the simple ladder structure studied here, we can hand design the update rule of state change.
For each species 1 particle (yellow), we check,
1). if this particle is direct contact with a species 0 particle, and
2). if there exists a species 2 particle as the exact diagonal neighbor.
The second condition again ensures the growth is in order to reduce potential errors described in Section~\ref{final_config}.
We show an example of a 40000-step forward simulation with a deterministic hand-designed rule in the accompanying supplementary video.



\section{Graph neural network as update rule}

\subsection{GNN architecture}
The specific architecture of the GNN we use in this paper is described in the main text. 
Regarding the parameters, the local information aggregation and update module is set to repeat only once.
The number of hidden neurons in each multi-layer perceptron are set to be 5.
The edge threshold for determining the connection of particles within the graph is set to be 3.0.
An additional bias of $-8.0$ is set for the last layer of the Sigmoid activation.


\subsection{Pretraining}
Since our system is a many-body dynamical system and state change is a low-probability event, the optimization landscape is characteristically flat and rugged.
To overcome this, we initiated the optimization process with a pretraining procedure. This involves mapping the state change probability determined by the GNN onto a probabilistic version of the hand designed rule. 
The loss function used in this phase is the L2 loss, calculated based on the differences in state change probabilities for each particle as predicted by the GNN and the hand-designed rule.
In practice, we originally attempted the state change mechanism with a set of standard neural networks, on which the pretraining phase for the GNN was subsequently based.

Fig. 2 in the main text showcases the use of the pretrained parameters as initial inputs. While these parameters do not yield ideal state change rule (as indicated by the red inset of Fig. 2 in the main text), they serve as a good starting point for subsequent optimization processes.

\subsection{Loss function}
As elaborated in the main text and reference \cite{jaxmd_patchy}, the optimization involves evaluating the loss of the final configurations and back-propagating it to update the parameters.
Consequently, the loss function is required to provide an accurate assessment of the parameterized update rule, and also be differentiable and conducive for navigating the optimization process, i.e. the loss should be smooth over the parameter space.

Considering our system under optimization, the loss function consists of two components. 
The first term quantifies the deviation of the assembled structures from the ideal structure of the same size.
This is achieved by calculating the number of particles that have undergone state change and constructing an ideal ladder structure of the same size.
We then take the L2 loss between the total potential energy of these two structures as the first term of the loss function. 
Here we particularly choose the L2 loss of energy, as opposed to the positional differences of each particle, because energy gives a smoother optimization landscape.

However, the first term alone will not be helpful for the optimization, as the system minimizes deviation simply by avoiding state change at all, resulting in a consistently zero loss. 
To mitigate this local minimum, a second term is introduced to encourage state change. 
This term is directly proportional to the number of particles that have undergone state change, with a coefficient of $6000$.

To maintain an effective balance these two terms, we set a saturation threshold of 5 for the second term, i.e. the total loss can be written as: Total loss = (energy difference)$^2$ - 
$6000\times\max(\text{Number of state changed particles}, 5)$.

\subsection{Optimization results}
The loss trajectory during the optimization process is presented in the main text Fig. 2(b). 
Here we provide in the accompanying supplementary video an example of a 100000-step forward simulation with the GNN update rule parameterized by the final optimized parameters.
This set of parameters are used in all subsequent studies presented in this paper.

\subsection{Yield curve of different GNN rules}
\begin{figure}
    \centering
    \includegraphics[width = 0.6\linewidth]{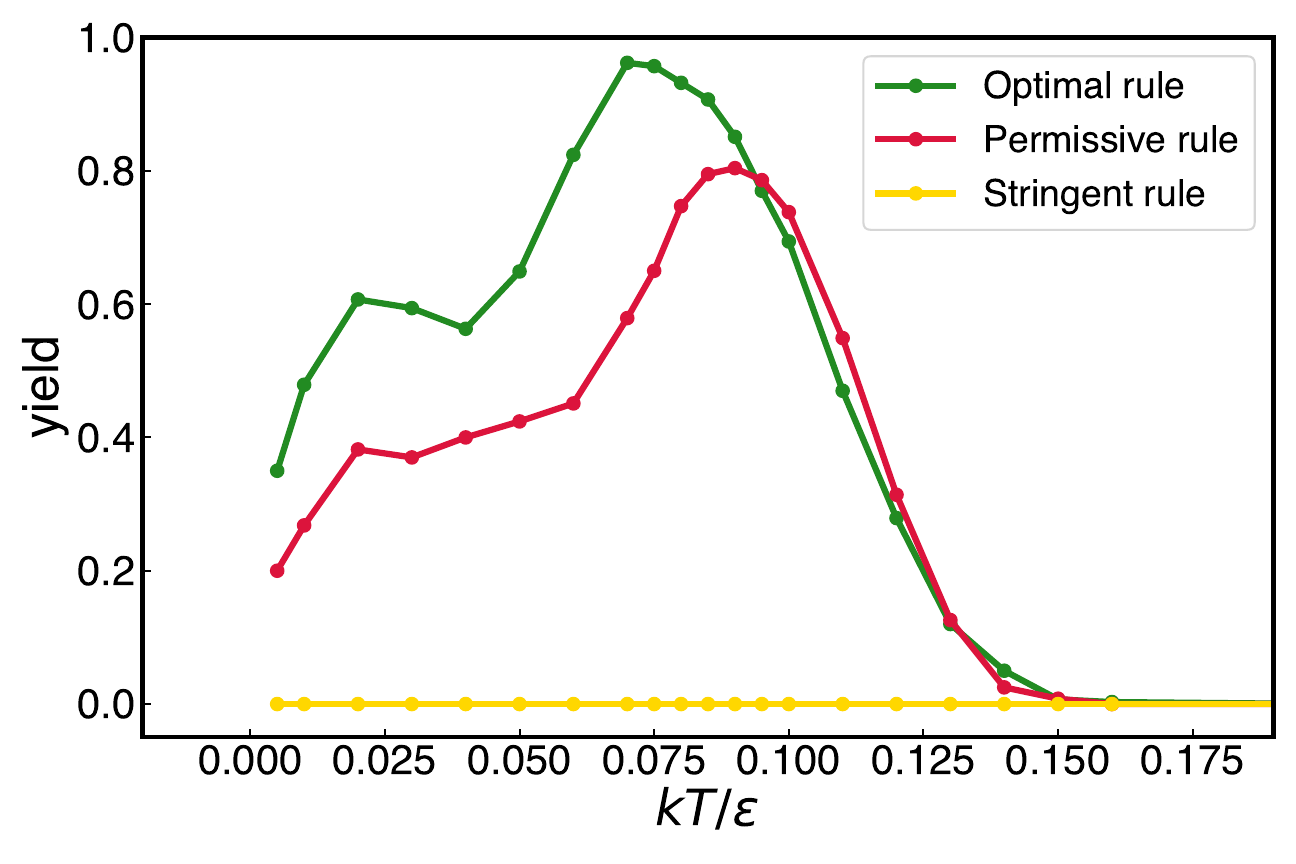}
    \caption{Yield curves of the optimal, over-permissive and over-stringent GNN rules respectively.}
    \label{fig:s3}
\end{figure}

In this section we present two examples that illustrate how yield curves deteriorate when the GNN rules are either over-permissive or over-stringent.
For the over-permissive rule, we choose the input parameters corresponding to the red inset of Fig. 2(b) in the main text. 
For the over-stringent rule, we choose the parameters at optimization step 50, corresponding to the yellow inset of Fig. 2(b) in the main text.
The optimal rule is represented by the final optimized parameters.
The yield curves were computed using the same definition in Section.~\ref{defyield}. 
The results are shown in Fig.~\ref{fig:s3}, demonstrating that significant yield increases depend solely on the optimal programming of the GNN rules.

\bibliographystyle{apsrev4-2}
\bibliography{si}